\documentclass[showpacs,preprintnumbers,amsmath,amssymb,prb,twocolumn,superscriptaddress,footinbib]{revtex4}
\usepackage{graphicx}
\usepackage{bm}
\usepackage{subfigure}
\newcommand{\bo}[1]{\mathbf{#1}}
\newcommand{\delH}[2]{\frac{\partial H}{\partial T^{#1}_{\bo{q}_{#2}}}}
\newcommand{\ddelH}[4]{\frac{\partial^2 H}{\partial T^{#1}_{\bo{q}_{#2}} \partial T^{#3}_{\bo{q}_{#4}}}}
\newcommand{\ex}[1]{\mathrm{e}^{#1}}
\newcommand{\im}{\mathrm{i}}
\newcommand{\sm}{\sum_{\{\bo{m}\}}}
\newcommand{\bz}{\, \mathrm{e}^{-H/T}}
\newcommand{\fv}[1]{\left\langle #1 \right\rangle}

\newcommand{\du}{\mathrm{d}}

\newcommand{\sr}{\sum_{\bo{r}}}
\newcommand{\Hx}[1]{\frac{\partial H}{\partial T^{x}_{\bo{q}_{#1}}}}
\newcommand{\Hxx}[2]{\frac{\partial^2 H}{\partial T^{x}_{\bo{q}_{#1}} \partial T^{x}_{\bo{q}_{#2}}}}

\begin{document}


\title{Instanton correlators and phase transitions in two- and three-dimensional logarithmic plasmas}


\author{K. B{\o}rkje}
\email[]{E-mail: kjetil.borkje(a)phys.ntnu.no}
\author{S. Kragset} 
\email[]{E-mail: steinar.kragset(a)phys.ntnu.no}
\author{A. Sudb{\o}}
\email[]{E-mail: asle.sudbo(a)phys.ntnu.no}
\affiliation{Department of Physics, Norwegian University of Science
  and Technology, N-7491 Trondheim, Norway}


\date{\today}

\begin{abstract}
  The existence of a discontinuity in the inverse dielectric constant
  of the two-dimensional Coulomb gas is demonstrated on purely
  numerical grounds. This is done by expanding the free energy in an
  applied twist and performing a finite-size scaling analysis of the
  coefficients of higher-order terms. The phase transition, driven by
  unbinding of dipoles, corresponds to the Kosterlitz-Thouless
  transition in the 2D XY model. The method developed is also used for
  investigating the possibility of a Kosterlitz-Thouless phase
  transition in a three-dimensional system of point charges
  interacting with a logarithmic pair-potential, a system related to
  effective theories of low-dimensional strongly correlated systems.
  We also contrast the finite-size scaling of the fluctuations of the
  dipole moments of the two-dimensional Coulomb gas and the
  three-dimensional logarithmic system to those of the
  three-dimensional Coulomb gas.
\end{abstract}


\maketitle


\section{\label{sec:intro} INTRODUCTION}

Compact $U(1)$ gauge fields in three dimensions are of great interest
in condensed matter theory, as they arise in effective theories of
strongly correlated two-dimensional systems at zero temperature.
\cite{marston,anderson1,anderson2,dagotto} Lightly doped Mott-Hubbard
insulators, such as high-$T_c$ cuprates, are examples of systems possibly
described by such theories, where the compact gauge field emerges from
strong local constraints on the electron dynamics.\cite{ioffe,
  anderson1,nagaosa,mudry} High-$T_c$ cuprates appear to fall outside
the Landau Fermi liquid paradigm, and a so-called
confinement-deconfinement transition in the gauge theories may be associated 
with breakdown of Fermi-liquid and quasiparticles in
2D at $T=0$.\cite{ichinose, nagaosa, mudry} Obliteration of
electron-like quasiparticles and spin-charge separation in the
presence of interactions is well known to occur in one spatial
dimension. However, the mechanism operative in that case, namely
singular forward scattering, is unlikely to be operative in higher
dimensions due to the much less restrictive kinematics at the Fermi
surface. \cite{leechen} Proliferation of instantons of emergent gauge
fields show more promise as a viable candidate mechanism. This line of
pursuit has recently been reinvigorated in the context of
understanding the physics of lightly doped Mott-Hubbard insulators and
unconventional insulating states.\cite{senthil-fisher}

The compact nature of a constraining gauge field on a lattice model
introduces topological defects defined by surfaces where the field
jumps by $2 \pi$, forming a gas of instantons (or "monopoles") in
$2+1$ dimensions.\cite{polyakov} Considering the gauge sector only,
the interactions between these instantonic defects are the same as
between charges in a 3D Coulomb gas, {\it i.e.} $1/r$-interactions.
Such a gas is always in a metallic or plasma phase with a finite
screening length, \cite{polyakov,kosterlitz1977} and there is no 
phase transition between a metallic regime and an insulating regime.
However, in models where compact gauge fields are coupled to {\it
matter fields}, the interaction between the magnetic monopoles may
be modified by the emergence of an anomalous scaling dimension of the
gauge field due to critical matter-field fluctuations.\cite{herbut2} This is the case 
for the compact abelian Higgs model with matter fields in the fundamental 
representation.\cite{sudbo} 

In Refs. \onlinecite{sudbo}, it was shown that the introduction of a
matter field with the fundamental charge leads to an anomalous scaling
dimension in the gauge field propagator \cite{herbut2}. The effect is to 
alter the interaction potential between the magnetic monopoles from $1/r$ 
to $-\ln r$. The existence of a confinement-deconfinement transition in 
the gauge theory is thus related to whether a phase transition occurs in 
a 3D gas of point charges with logarithmic interactions. However, one 
should note that the legitimacy of a monopole action based on just pairwise 
interactions has been questioned, particularly when viewed as an effective
description of an effective gauge theory of strongly interacting 
systems. \cite{hermele} The 3D logarithmic plasma is however of considerable 
interest in its own right. 

In two dimensions, where $-\ln r$ is the
Coulomb potential, it is known that the logarithmic gas experiences a phase
transition from a low-temperature insulating phase consisting of
dipoles to a high-temperature metallic phase. This is nothing but the
Coulomb-gas representation of the Kosterlitz-Thouless transition in
the 2D XY model. In a 3D logarithmic gas, the existence of a phase transition
is still subject to debate.\cite{sudbo,herbut,chernodub}
Renormalization group arguments have been used \cite{sudbo} to
demonstrate that a transition may occur, driven by the unbinding of
dipoles. Others have claimed that the 3D logarithmic gas is always in the
metallic phase.\cite{herbut} In a recent paper,\cite{kragset} large
scale Monte Carlo simulations indicated that two distinct phases of
the 3D-log gas exists; a low-$T$ regime where the dipole moment does
not scale with system size and a high-$T$ regime where the dipole
moment is system size dependent. Those results do however not determine the character
of the phase transitions. That will be the main subject of this paper.

The Kosterlitz-Thouless transition in the 2D XY model is characterized
by the universal jump to zero of the helicity
modulus.\cite{kosterlitz} In the corresponding 2D Coulomb gas, it is
the inverse of the macroscopic dielectric constant $\epsilon$ that
experiences a jump to zero when going from the insulating to the
metallic phase. According to Ref. \onlinecite{sudbo}, such a universal
discontinuity should also take place for $\epsilon^{-1}$ in the 3D logarithmic
gas associated with the confinement-deconfinement transition. Proving that 
such discontinuities exist numerically is a subtle task. The discontinuous
character of the helicity modulus in the 2D XY model is very hard to
see in a convincing manner by computing the helicity modulus, due to
severe finite-size effects. It was only recently proven on purely
numerical grounds that such a discontinuity exists \cite{minnhagen} in
a simple, but yet clever manner. By imposing a twist across the system
and expanding the free energy in this twist to the fourth order, a
stability argument was used to show that the second order term in the
expansion, the helicity modulus, must be nonzero at $T_c$. The proof relies
on the ability to conclude that the fourth order term is negative in the
thermodynamic limit, from which the discontinuity follows immediately. In this
paper, we will repeat this procedure, but now in the language of the
2D Coulomb gas. In addition to confirming the results of Minnhagen and
Kim, the method which we develop here could be suitable for
proving the possibly discontinuous behaviour of $\epsilon^{-1}$ in the
3D logarithmic gas. This is a main motivation for translating the procedure of
Ref. \onlinecite{minnhagen} to the vortex language, since the 3D logarithmic
gas is not the dual theory of any simple spin model. After having demonstrated 
the discontinuity in the 2D Coulomb gas, we go on to apply the method on the
3D logarithmic gas. We also compare the scaling with system size of the
mean square dipole moment for these logarithmic plasmas, and contrast the results with those of
the 3D Coulomb gas. This is important, since the mean square dipole moment does not scale with system size
below a certain temperature for the logarithmic plasmas.\cite{kragset} This indicates that two phases exist, where the
low-temperature regime consists of tightly bound pairs. However, the results for the 3D Coulomb gas
are qualitatively different, in accordance with the fact that such a low-temperature phase is absent in that case.

\section{\label{sec:model} MODEL}

The Hamiltonian of the 2D XY model on a square lattice modified with a
twist $\bo{T}(x,y)$ is
\begin{equation}
  \label{eq:ham2DXY}
  H_{\mathrm{XY}} = -J \sum_{\langle i,j \rangle} \cos (\theta_i - \theta_j - 2 \pi \, \bo{r}_{ij} \cdot \bo{T}),
\end{equation} 
where $\bo{r}_{ij}$ is the displacement between the nearest neighbour
pairs to be summed over. We set the coupling constant $J$ to unity.
The volume of the system, {\it i.e.} the number of lattice points, is
$L^2$, and the angle $\theta_{i}$ is subject to periodic boundary
conditions. In the Villain approximation, a duality transformation
leads to the Hamiltonian
\begin{equation}
  \label{eq:ham2DCG}
  H = \frac{1}{2} \sum_{i,j} (m + \varepsilon^{\mu \nu} \Delta^{\mu} T^{\nu})_i V_{ij} ( m + \varepsilon^{\rho \sigma} \Delta^{\rho} T^{\sigma})_j,
\end{equation}
where $m_i$ are point charges on the dual lattice, corresponding to
vortex excitations in the XY model. $\Delta^{\mu}$ is a lattice
derivative and $\varepsilon^{\mu \nu}$ is the completely
anti-symmetric symbol. The potential $V_{ij}$ is given by
\begin{equation}
  \label{eq:pot2DCG}
  V(|\bo{r}_i - \bo{r}_j|) = \frac{2 \pi^2}{L^2} \sum_{\bo{q}} \frac{\ex{- \im \bo{q} \cdot (\bo{r}_i -\bo{r}_j)}}{2-\cos q_x - \cos q_y},
\end{equation}
which has a logarithmic long-range behaviour. Details of the
dualization are found in appendix \ref{sec:appdual}.  As is well
known, eq. \eqref{eq:ham2DCG} at zero twist describes the
two-dimensional Coulomb gas (2D CG). In this representation, the
Kosterlitz-Thouless phase transition of the 2D XY model is recognized
by a discontinuous jump to zero of the inverse macroscopic dielectric
constant $\epsilon^{-1}$ at $T_c$. We note that the curl of the twist $\bo{T}$ acts as a modification of the charge field in the 2D CG.

The free energy of the system is $F = -T \ln Z$, where the partition
function is given by summing the Boltzmann factor over all charge
configurations:
\begin{equation}
  \label{eq:partfunction}
  Z = \sum_{\{m\}} \ex{-H/T}.
\end{equation}
Let us write the Hamiltonian in Fourier representation,
\begin{equation}
  \label{eq:ham2DCGFourier}
  H = \frac{1}{2 L^2} \sum_{\bo{q}} \left(m_{\bo{q}} + \varepsilon^{\nu \lambda} Q^{\nu}_{-\bo{q}} T^{\lambda}_{\bo{q}} \right)V_{\bo{q}} \left(m_{-\bo{q}} + \varepsilon^{\rho \sigma} Q^{\rho}_{\bo{q}} T^{\sigma}_{-\bo{q}} \right),
\end{equation}
where the discrete Fourier transform is defined as in appendix
\ref{sec:appTaylor} and $\Delta^{\mu} \ex{\pm \im \bo{q} \cdot \bo{r}}
\equiv \ex{\pm \im \bo{q} \cdot \bo{r}} Q^{\mu}_{\pm \bo{q}}$.

\section{\label{sec:stability} STABILITY ARGUMENT}

From \eqref{eq:ham2DXY}, it is clear that $F(\bo{T}) \geq F(0)$ in the
low-temperature phase, {\it i.e.} the free energy is minimal for zero
twist. This inequality is also valid at the critical temperature
$T_c$, since the free energy must be a continuous function of
temperature. As a consequence, the Taylor expansion
\begin{equation}
  \label{eq:TaylorF}
  \begin{split}
  F(\bo{T}) - F(0) & = \sum_{\alpha} \sum_{\bo{q}_1} \frac{\partial F}{\partial T^{\alpha}_{\bo{q}_1}}\bigg|_{\bo{T} = 0} T^{\alpha}_{\bo{q}_1} \\
& + \sum_{\alpha, \beta} \sum_{\bo{q}_1 \bo{q}_2} \frac{\partial^2 F}{\partial T^{\alpha}_{\bo{q}_1} \partial T^{\beta}_{\bo{q}_2}}\bigg|_{\bo{T} = 0} \frac{T^{\alpha}_{\bo{q}_1} T^{\beta}_{\bo{q}_2}}{2} + ...
\end{split}
\raisetag{1cm}
\end{equation}
can not be negative for any $T \leq T_c$. Expressions for the
derivatives of the free energy with respect to a general twist are
found in appendix \ref{sec:appTaylor}. Only terms of even order will
contribute to the series, since $m_i$ may take equally many positive
and negative values. We are free to {\it choose} the twist to be
\begin{equation}
  \label{eq:twistchoice}
  \bo{T}(x,y) = \frac{\Delta}{L^\eta} \sin \left(\frac{2 \pi \, y}{L}\right) \bo{\hat{x}},
\end{equation}
where $\Delta$ is an arbitrarily small constant and $\eta = 1$ for the two-dimensional Coulomb gas. To the fourth order,
this long-wavelength twist turns \eqref{eq:TaylorF} into
\begin{equation}
  \label{eq:finalexpr}
  \begin{split}
   & F(\bo{T}) - F(0) = \frac{\Delta^2}{4} C_{\bo{k}} \left(1-\frac{V_{\bo{k}}}{L^2 T} \fv{m_{\bo{k}} m_{-\bo{k}}} \right) \\
& + \frac{\Delta^4}{32} \frac{(C_{\bo{k}} V_{\bo{k}})^2}{L^4 T^3} \left(\fv{m_{\bo{k}} m_{-\bo{k}}}^2 - \frac{1}{2} \fv{(m_{\bo{k}} m_{-\bo{k}})^2}\right),
\raisetag{1cm}
\end{split}
\end{equation}
where $\bo{k} = (0,2 \pi/L)$ and $C_{\bo{k}} = Q^y_{\bo{k}}
Q^y_{-\bo{k}} V_{\bo{k}}$. We recognize the paranthesis in the second
order term as the dielectric response function $\epsilon^{-1}(\bo{k})$, where $\bo{k}$ is now the
smallest nonzero wave vector in a finite system. Note that the
prefactors in both terms are independent of system size as $L
\rightarrow \infty$. The crucial argument to use is the same as in
Ref. \onlinecite{minnhagen}. If the fourth order term approaches a
finite negative value at $T_c$ in the limit $L \rightarrow \infty$,
the second order term, $\epsilon^{-1}(\bo{k} \rightarrow 0)$, must be
positive to satisfy the inequality $F(\bo{T}) \geq F(0)$. Furthermore,
since we know that the inverse dielectric constant is zero in the
high-temperature phase, it necessarily experiences a discontinuity at
$T_c$. As we shall see, Monte Carlo simulations show that the fourth
order term is indeed negative at $T_c$ in the thermodynamic limit.

The argument described above will also apply to a three-dimensional gas
of point charges interacting via a pair potential of some sort, as
long as the twist raises the free energy in the low-temperature
regime. Since the curl of the twist $\bo{T}$ is a vector in that case, 
one may for instance choose the $z$-component of this vector as the
perturbing charge in eq. \eqref{eq:ham2DCG}. The two three-dimensional
systems we will consider are the logarithmic gas and the Coulomb
gas. The expansion \eqref{eq:TaylorF} is valid for any system size
$L$. However, to make the change in free energy nondivergent as $L
\rightarrow \infty$, the twist must be chosen such that the terms in
the expansion are independent of system size. This is obtained by
choosing $\eta=2$ for the logarithmic gas and $\eta=3/2$ for the
Coulomb gas. $\eta$ is defined in \eqref{eq:twistchoice}.  
In both cases, the second order term will be proportional to 
\begin{equation}
  \label{eq:invdielec3D}
  \epsilon^{-1}(\bo{k}) = 1-\frac{V_{\bo{k}}}{L^3 T} \fv{m_{\bo{k}} m_{-\bo{k}}}.
\end{equation}
The fourth order term will be proportional to
\begin{equation}
  \label{eq:fourthorder3D}
  \epsilon_4(\bo{k}) \equiv \frac{1}{T^3} \left(\fv{m_{\bo{k}} m_{-\bo{k}}}^2 - \frac{1}{2} \fv{(m_{\bo{k}} m_{-\bo{k}})^2}\right)
\end{equation}
in the logarithmic case. In the case of a 3D Coulomb gas, the interesting quantity will be $\epsilon_4/L^2$, 
which is independent of system size since $\fv{m_{\bo{k}} m_{-\bo{k}}} \sim L$ in that case.

\section{\label{sec:simulation} SIMULATION RESULTS}

Standard Metropolis Monte Carlo simulations are carried out on the
model \eqref{eq:ham2DCG} at zero twist. An $L \times L$
square lattice with periodic boundary conditions is used and the
system is kept electrically neutral at all times during the
simulations. This is achieved by inserting dipoles with probability
according to the Metropolis algorithm: An insertion of a negative or
positive charge is attempted at random at a given lattice site, and an
opposite charge is placed at one of the nearest neighbour sites to
make the dipole. This is one move, accepted with probability $\exp
(-\Delta E / T) = \exp [-(\mathcal{H_{\mathrm{new}}} -
\mathcal{H_{\mathrm{old}}})/T]$, and the sequence of trying this for
all sites in the system once is defined as one sweep. If a charge is
placed on top of an opposite one, the effect is to annihilate the
existing one. All simulations are performed going from high to low
temperature and after simulating one system size $L$ the sampled data
are postprocessed using Ferrenberg-Swendsen reweighting
techniques.\cite{ferr1}

\subsection{\label{subsec:sim2DCD} 2D Coulomb gas} 
We consider first the 2D Coulomb gas, which is known to suffer a
metal-insulator transition via a Kosterlitz-Thouless phase transition.
In this case, Monte Carlo data are obtained for $L = 4- 100$ and for
each $L$ up to 200 000 sweeps at each temperature is used.

We start by taking the Hamiltonian (\ref{eq:ham2DCG}) and computing
the mean square of the dipole moment, $\langle s^2 \rangle$, as a
function of system size and temperature. A mean square
dipole moment which is independent of system size indicates the
existence of tightly bound dipoles and a dielectric or insulating
phase. If the mean square dipole moment scales with system size, this
demonstrates the existence of free unbound charges and hence a
metallic phase. In other words, we expect in the low-temperature
dielectric insulating phase no finite-size scaling of
$\langle s^2 \rangle$, whereas we should expect $\langle s^2 \rangle
\propto L^{\alpha (T)}$ with $\alpha (T) \leq 2$ at higher
temperatures. Using an intuitive low density argument, neglecting
screening effects,\cite{KosterlitzThouless} we can calculate the
behaviour of $\langle s^2 \rangle$ to leading order in $L$, 
\begin{equation}
  \label{eq:ssquared}
  \langle s^2 \rangle \propto \begin{cases} {\rm Const.} & ; \; \; T < T_{KT} \\
    L^{ (T - T_{KT})/T } & ; \; \; T_{KT} < T < 2 T_{KT} \\
    L^2 & ;  \; \; 2 T_{KT} < T.
    \end{cases}
\end{equation}
Hence, $\alpha(T)$ is zero for low temperatures and a monotonically
increasing function of temperature just above $T_{KT}$. Including
screening effects in 2D shows that this conclusion still holds,
however the temperature at which it occurs is determined by screening.

\begin{figure}[htbp]
  \centering
  \scalebox{0.7}{
    \hspace{-1cm}
    \includegraphics{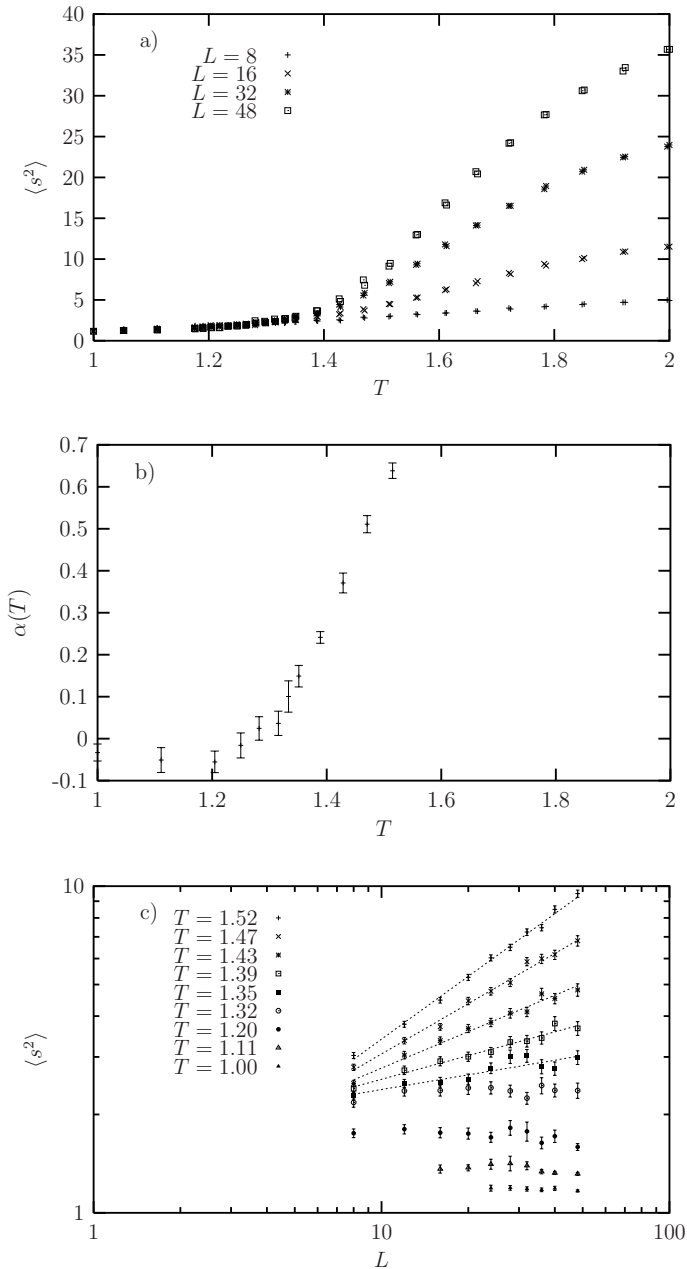}
  }
  \caption{The mean square dipole moment $\langle s^2 \rangle$ as a
    function of temperature (top panel), and system size (bottom
    panel) for the 2D Coulomb gas. The middle panel shows the scaling
    exponent $\alpha$ extracted from $ \langle s^2 \rangle \sim
    L^{\alpha(T)}$. }
\label{fig:MCData2d.eps}
\end{figure}

Details of the simulations may be found in Ref. \onlinecite{kragset}.
The result is shown in Fig. \ref{fig:MCData2d.eps} where we have the
mean square dipole moment for the 2D case both as a function of
temperature for various system sizes, and as function of system size
for various temperatures. From this we may extract the scaling
constant $\alpha(T)$ which is shown in the center panel of Fig.
\ref{fig:MCData2d.eps}. A related method for using dipole fluctuations
to measure vortex-unbinding has recently been used in Refs.
\onlinecite{fertig}.

Below a temperature $ T \approx 1.3$, no scaling of $ \langle s^2
\rangle$ is seen, consistent with a low-temperature dielectric phase.
The temperature at which scaling stops is consistent with the known
temperature at which the 2D Coulomb gas suffers a metal-insulator
transition.

\begin{figure}[htbp]
  \centering
  \scalebox{0.8}{
    \hspace{-1cm}
    \includegraphics{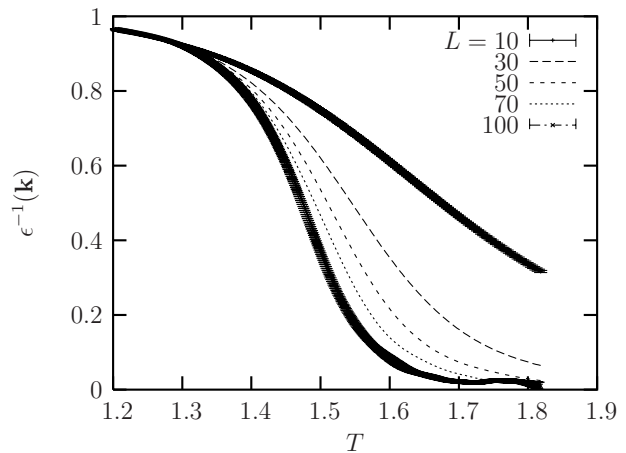}
  }
  \caption{Inverse dielectric constant taken at the smallest possible wave
    vector in a finite system, $\mathbf{k} = (0,2 \pi / L)$, and
    plotted against temperature $T$ for system sizes $L =$ 10, 30, 50,
    70 and 100, for the 2D Coulomb gas. The decrease of
    $\epsilon^{-1}$ towards zero becomes sharper with increasing $L$,
    consistent with the prediction of a discontinuous jump. Errorbars
    are given in the top and bottom curves, and omitted for clarity in
    the others.}
  \label{fig:InvDielec2D}
\end{figure}

\begin{figure}[htbp]
  \centering
  \scalebox{0.8}{
    \hspace{-0.5cm}
    \includegraphics{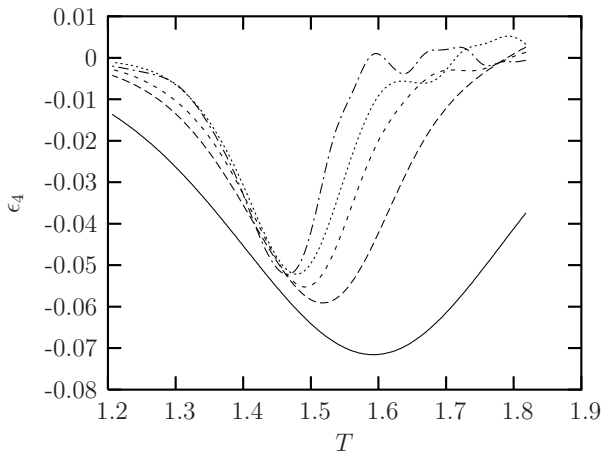}
  }
  \caption{The coefficient $\epsilon_4$ of the fourth order term  of
    the expansion of the free energy, for the 2D Coulomb gas. The same
    systems are used in this plot as in Figure \ref{fig:InvDielec2D},
    and the depths decrease with increasing $L$. The important
    question is whether this dip vanishes at $T_c$ or not. Errorbars
    are omitted but will be reintroduced in Figure \ref{fig:Depth2D}.
    The oscillation at high $T$ is due to noise from the reweighting.}
  \label{fig:FourthOrder2D}
\end{figure}

Simulation results for the inverse dielectric constant are shown for a
selection of system sizes in Fig. \ref{fig:InvDielec2D}. Since
$\epsilon^{-1}$ is expected to be discontinuous at $T_c$ in the limit
$\mathbf{k} \rightarrow 0$, we consider only the smallest possible
wave vector in each system, $\mathbf{k} = (0,2 \pi / L)$, and we see
that the decrease of $\epsilon^{-1}$ towards zero with increasing $T$
indeed gets sharper as $L$ grows. It is however difficult to decide
from these plots alone whether or not the dielectric constant is
discontinuous at $T_c$. The fourth order term in the expansion of the
free energy, $\epsilon_4$ defined in eq. \eqref{eq:fourthorder3D}, is therefore 
investigated in a corresponding manner and plotted in
Fig. \ref{fig:FourthOrder2D}.

\begin{figure}[htbp]
  \centering \scalebox{0.8}{ 
    \hspace{-2cm}
    \includegraphics{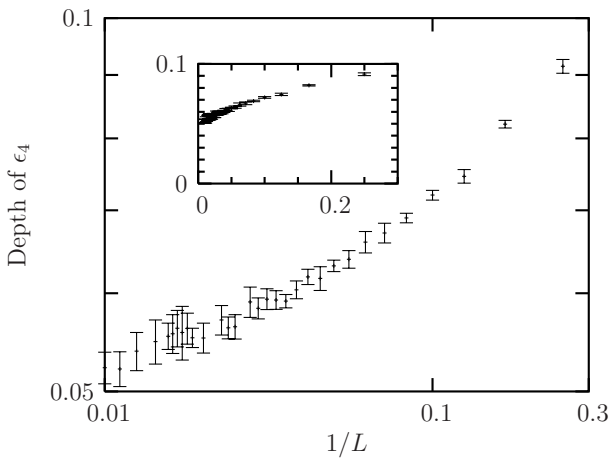} }
  \caption{Depth of the dip in the fourth order term shown in Figure
    \ref{fig:FourthOrder2D} for the 2D Coulomb gas. The data are
    obtained from simulations of system sizes ranging from $L = 4$ to
    $L = 100$ and plotted both on a linear scale (inset) and on a
    log-log scale. The positive curvature in the log-log plot clearly
    indicates a nonzero value of the depth when extrapolating to the
    limit $L \rightarrow \infty$.}
  \label{fig:Depth2D}
\end{figure}

We note that this quantity has a dip at a temperature which can be
associated with the transition temperature. If this dip remains finite
and negative as $L$ approaches infinity, $\epsilon^{-1}$ must exhibit
a jump at $T_c$. The depth of the dip is shown in Fig.
\ref{fig:Depth2D} for a variety of system sizes ranging from $L=4$ to
$L=100$ and as a function of $1/L$. It clearly decreases with
increasing $L$. However, from the positive curvature of the data in
the log-log plot we may conclude that the depth remains nonzero when
we extrapolate to $1/L = 0$, a conclusion reached by assuming
power-law dependence of the depth on $L$.

We can now subtract from the depth a constant chosen so as to
linearize the curve in the log-log plot. This constant consequently
corresponds to the depth when extrapolating the data to the
thermodynamic limit $1/L = 0$, and we find this to be $0.047 \pm
0.005$.

\begin{figure}[htbp]
  \centering \scalebox{0.8}{ 
    \hspace{-2cm}
    \includegraphics{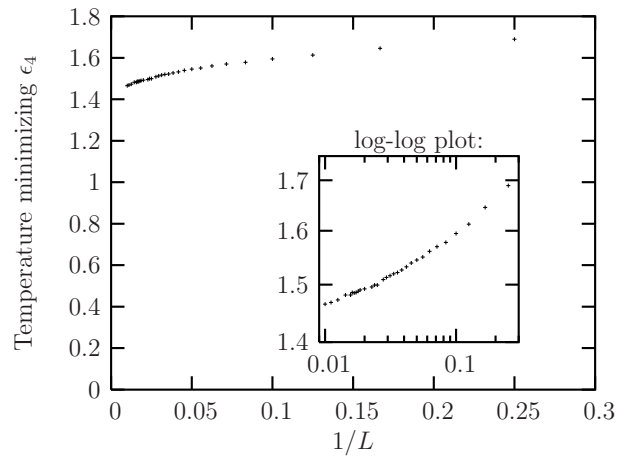} }
  \caption{Temperature minimizing $\epsilon_4$ as a function of
    inverse system size for the 2D CG. The values are plotted both on
    a linear scale and on a log-log scale (inset). This temperature
    reaches a nonzero value at $L \rightarrow \infty$ indicated by the
    positive curvature in the log-log plot. Extrapolation gives $T_c =
    1.36 \pm 0.04$.}
  \label{fig:Tc2D}
\end{figure}

By plotting the temperature at which the fourth order term has its
minimum against $1/L$, we can follow a similar procedure as the above
one. This is shown in Fig. \ref{fig:Tc2D}.  We linearize a log-log
plot by subtracting a carefully chosen constant and end up with the
number $1.36 \pm 0.04$. This is nothing else than an estimate of the
critical temperature of the 2D CG, and compares well to earlier
results.\cite{olssonpers} The approach towards $T_c$ is however a bit
slow, making a precise determination of the critical temperature
difficult.  This drawback was also noted by Minnhagen and Kim for the
corresponding computations on the 2D XY model.\cite{minnhagen}

\subsection{\label{subsec:sim3DLG} 3D logarithmic system}
\begin{figure}[htbp]
  \centering
  \scalebox{0.7}{
    \hspace{-1cm}
    \includegraphics{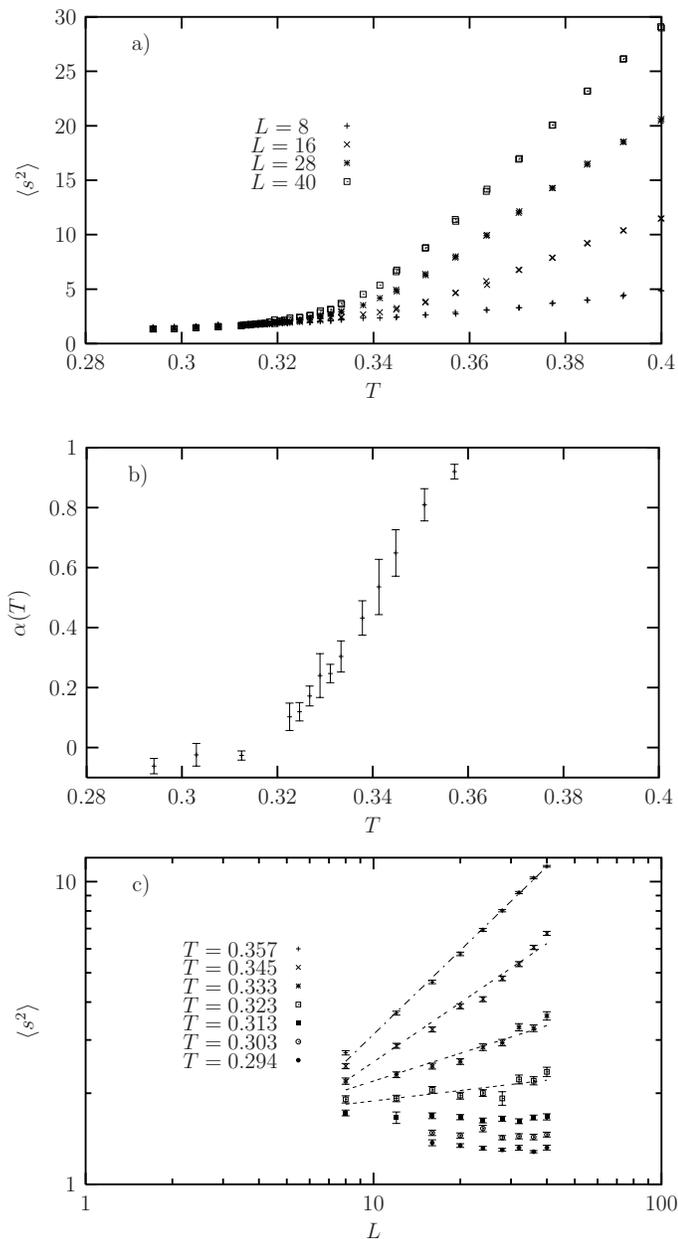}
  }
\caption{Mean square dipole moment $\langle s^2 \rangle$ as a function
  of temperature (top panel) and system size (bottom panel) for the 3D
  system of point charges interacting with a logarithmic bare pair
  potential (3D LG).  The middle panel shows the scaling exponent
  $\alpha$ extracted from $ \langle s^2 \rangle \sim L^{\alpha(T)}$.
}
\label{fig:MCData3d.eps}
\end{figure}

We may carry out the same type of analysis for the mean square dipole
moment for a system of point charges interacting via a
three-dimensional logarithmic bare pair potential (3D LG).
For this system, much less is known. Such a system has recently been
considered in the context of studying confinement-deconfinement phase
transitions in the $(2+1)$-dimensional abelian Higgs model.\cite{sudbo}
The results are shown in Fig. \ref{fig:MCData3d.eps}.

Qualitatively and quantitatively the results are the same in the 3D LG
as for the 2D case. This strongly suggests that the 3D LG also has a
low-temperature dielectric insulating phase separated by a phase
transition from a high-temperature phase. In the low-temperature
regime the charges of almost all dipoles are bound as tightly as
possible, the separation of the charges correspond to the lattice
constant. In the high-temperature regime the dipoles have started to
separate, reflected by a scaling of $\langle s^2 \rangle \sim
L^{\alpha (T)}$ with the system size. Since $\alpha (T) = 0$ at low
temperatures while $\alpha (T) \neq 0$ in the high-temperature regime
a non-analytic behaviour of $\alpha (T)$ is implied. This necessarily
corresponds to a phase transition in the vicinity of $T \approx 0.3$,
a temperature which agrees well with Ref. \onlinecite{sudbo} where a
critical value of $T_c = 1/3$ was obtained.

Note that, although this
simple type of analysis of the mean square dipole moment does not by itself suffice to determine the
character of these phase transitions either in the case of 3D LG or 2D CG, it
does suffice to shed light on the important issue of whether a low
temperature insulating phase exists in the 3D LG as well. This is far from
obvious, since the screening properties of a three-dimensional system
of charges interacting logarithmically is quite different from that of
a Coulomb system (in any dimension).\cite{herbut} It is therefore of
considerable interest to repeat the analysis carried out for the 2D
Coulomb gas to, if possible, determine the character of a
metal-insulator transition in the 3D LG.

\begin{figure}[htbp]
  \centering
  \scalebox{0.8}{
    \hspace{-1cm}
    \includegraphics{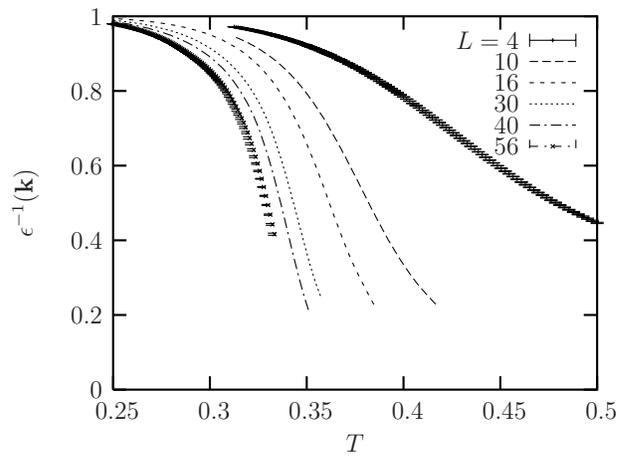}
  }
  \caption{Inverse dielectric constant taken at the smallest possible wave
    vector in a finite system, $\mathbf{k} = (0,2 \pi / L, 0)$, and
    plotted against temperature $T$ for system sizes $L =$ 4, 10, 16,
    30, 40 and 56, for the 3D LG system. The decrease of $\epsilon^{-1}$
    towards zero becomes sharper with increasing $L$, consistent with
    the prediction of a discontinuous jump. However, the downward
    drift in the temperature at which the inverse dielectric constant
    starts decreasing rapidly is more pronounced than in the 2D CG case.
    Errorbars are given in the top and bottom curves, and omitted for
    clarity in the others. }
  \label{fig:InvDielec3D}
\end{figure}

In Fig. \ref{fig:InvDielec3D} we show the inverse dielectric
constant for the 3D LG as a function of temperature for various system
sizes. It shows qualitatively the same behavior as for the 2D CG in
that the decrease of $\epsilon^{-1}$ towards zero becomes sharper with
increasing $L$. However, the downward drift in the temperature at
which the inverse dielectric constant starts decreasing rapidly is more
pronounced than in the 2D CG case.

\begin{figure}[htbp]
  \centering
  \scalebox{0.8}{
    \hspace{-0.5cm}
    \includegraphics{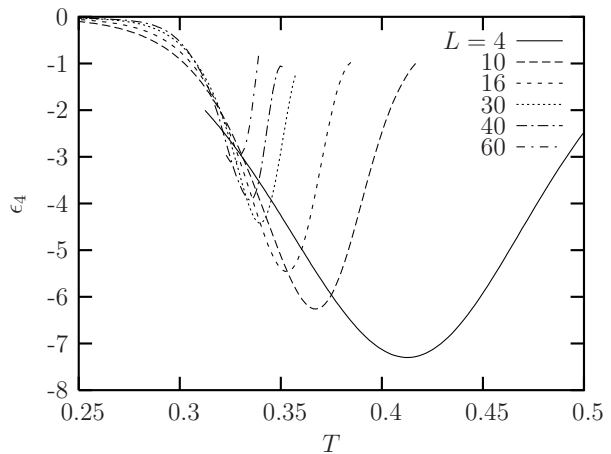}
  }
  \caption{The coefficient $\epsilon_4$ of the fourth order term of
    the expansion of the free energy for the 3D LG model. The depths
    decrease with increasing $L$, and the important question is
    whether this dip vanishes at $T_c$ or not. Errorbars are omitted
    but will be reintroduced in Figure \ref{fig:Depth3D}. }
  \label{fig:FourthOrder3D}
\end{figure}

\begin{figure}[htbp]
  \centering \scalebox{0.8}{ 
    \hspace{-2cm}
    \includegraphics{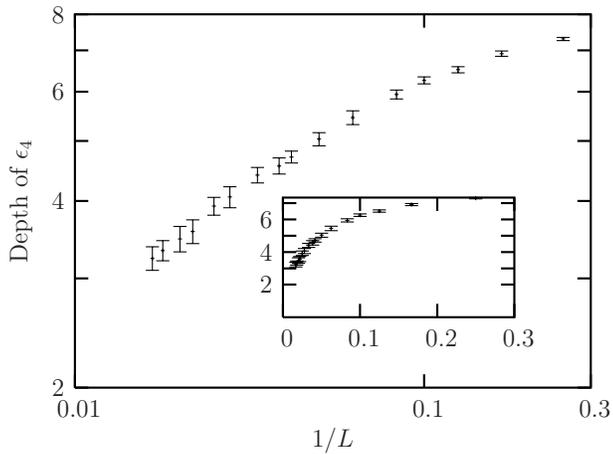} }
  \caption{Depth of the dip in the fourth order term shown in Figure
    \ref{fig:FourthOrder3D} for the 3D LG. The data are obtained from
    simulations of system sizes ranging from $L = 4$ to $L = 60$ and
    plotted both on a linear scale (inset) and on a log-log scale. The
    lack of clear positive curvature in the log-log plot that was
    observed in 2D CG case makes the extrapolation to the limit $L
    \rightarrow \infty$ more difficult for the system sizes we have
    been able to access in 3D.}
  \label{fig:Depth3D}
\end{figure}

In Fig. \ref{fig:FourthOrder3D} we have plotted the fourth order
coefficient against temperature for the 3D LG system, and 
the depth of the dip as a function of system size is shown in Fig. \ref{fig:Depth3D}.
It would clearly have been
desirable to be able to access larger system sizes than what we have
been able to do in the 3D LG case, to bring out a potential positive curvature
that was observed in the 2D CG case. From these results, it is unfortunately not possible to tell whether 
the depth of the dip remains finite and negative as $L \to
\infty$ or if it vanishes. Hence, we are presently not able to firmly 
conclude that the inverse dielectric constant in the 3D LG experiences a discontinuity.

\begin{figure}[htbp]
  \centering \scalebox{0.8}{ 
    \hspace{-2cm}
    \includegraphics{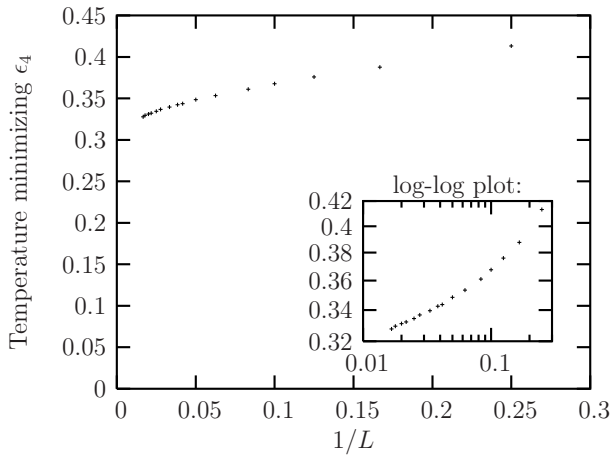} }
  \caption{Temperature minimizing $\epsilon_4$ as a function of
    inverse system size for the 3D LG system. The values are plotted
    both on a linear scale and on a log-log scale (inset). This
    temperature reaches a nonzero value at $L \rightarrow \infty$.
    Extrapolation gives $T_c = 0.30 \pm 0.04$.}
  \label{fig:Tc3D}
\end{figure}

The temperature locating the minimum in $\epsilon_4$ as a function of
system size is shown in Fig. \ref{fig:Tc3D} for the 3D LG system. Extrapolation gives $T_c = 0.30 \pm 0.04$.

\subsection{3D Coulomb gas}

\begin{figure}[htbp]
  \centering
  \scalebox{0.7}{
    \hspace{-2cm}
    \includegraphics{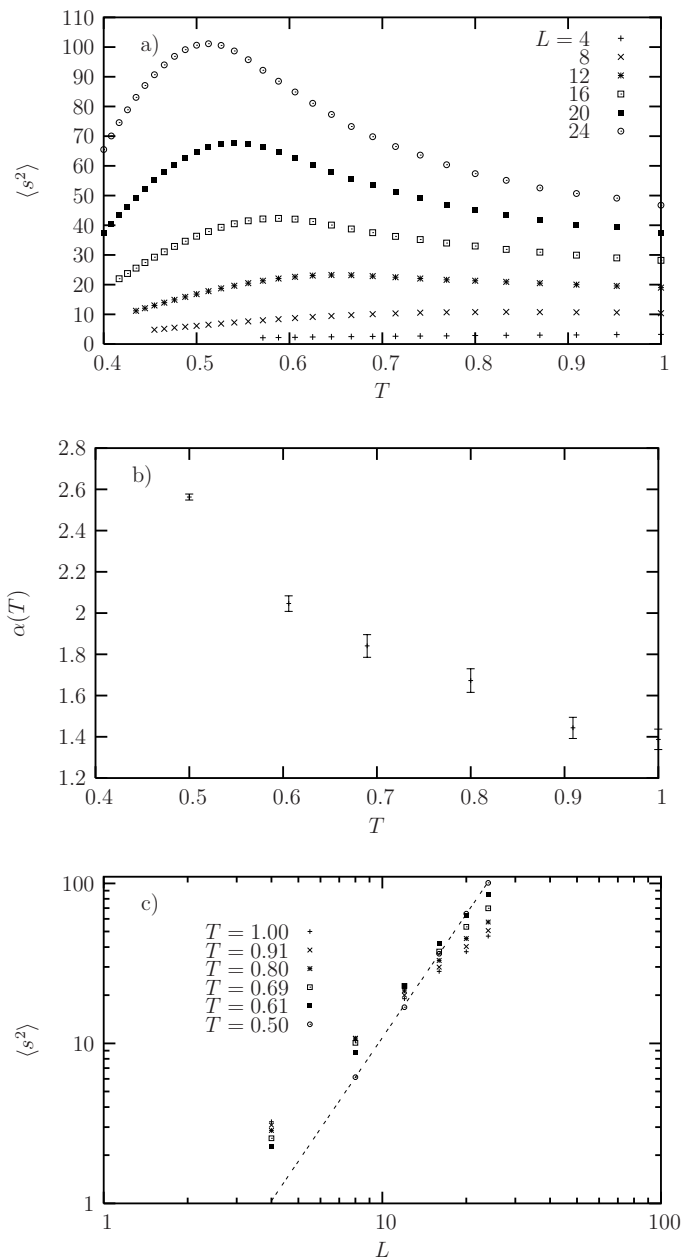}
  }
\caption{Mean square dipole moment $\langle s^2 \rangle$ as a function
  of temperature (top panel) and system size (bottom panel) for the 3D
  Coulomb gas system of point charges interacting with a $1/r$
  bare pair potential.  The middle panel shows the scaling exponent
  $\alpha$ extracted from $ \langle s^2 \rangle \sim L^{\alpha(T)}$.
}
\label{fig:MCData3dC.eps}
\end{figure}

In this subsection, we contrast the results of the 2D Coulomb gas and
the 3D LG to those of the 3D Coulomb gas. The 3D CG is known to be in a metallic
high-temperature phase for {\it all} finite temperatures and should
exhibit quite different finite-size scaling of $\langle s^2 \rangle$
compared to the 2D CG case.\cite{polyakov,kosterlitz77,sudbo} The
results are shown in Fig. \ref{fig:MCData3dC.eps}.  Note that the
temperature dependence of the curves for all different system sizes
are qualitatively different in the 3D CG compared to those in the 2D CG
and the 3D LG. This becomes particularly apparent upon considering the
$L$-dependence of $\langle s^2 \rangle$ for various temperatures, where the
steepness of the curves increases with decreasing temperature,
resulting in a scaling exponent $\alpha(T)$ (from $\langle s^2 \rangle \sim
L^{\alpha(T)}$) which decreases with increasing temperature.  This is
quite consistent with what is known for the 3D CG, namely that it exhibits a
metallic state for all finite temperatures, equivalently it
corresponds to Polyakov's permanent confinement.\cite{polyakov,sudbo}
It is evident that the scaling results for $ \langle s^2 \rangle $ for
the 2D CG and the 3D LG are qualitatively and quantitatively the same,
and that they are qualitatively different from those exhibited by the
3D CG. For low temperatures, $\langle s^2 \rangle$ seem to be increasing with temperature.
This is only a vacuum effect, since vacuum configurations do not contribute to the 
measurement of $\langle s^2 \rangle$ \cite{kragset}.
This means that close to vacuum, only configurations resulting from the insertion of one single
dipole at the smallest possible distance will contribute. See also section \ref{subsec:ladning}.

\begin{figure}[htbp]
  \centering
  \scalebox{0.8}{
    \hspace{-1cm}
    \includegraphics{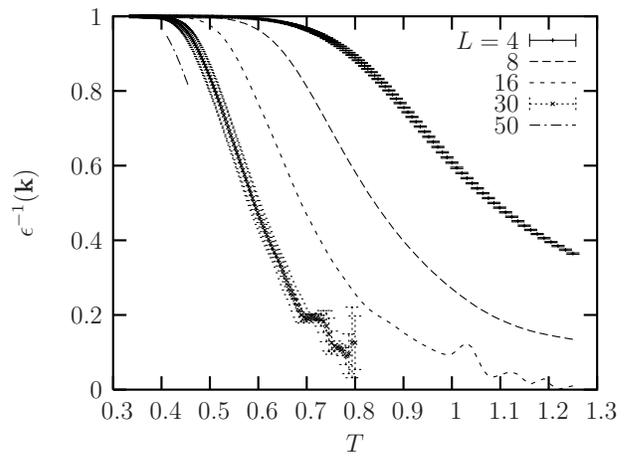}
  }
  \caption{Inverse dielectric constant taken at the smallest possible wave
    vector in a finite system, $\mathbf{k} = (0,2 \pi / L, 0)$, and
    plotted against temperature $T$ for system sizes $L =$ 4, 8, 16,
    30 and 50, for the 3D CG system. The decrease of $\epsilon^{-1}$
    towards zero does not sharpen with increasing $L$, and there is a
    clear downward drift in the temperature at which $\epsilon^{-1}$
    deviates from unity. Errorbars are given in two of the curves, and
    omitted for clarity in the others. }
  \label{fig:InvDielec3Dc}
\end{figure}

The inverse dielectric constant for the 3D CG is shown as a function
of temperature in Figure \ref{fig:InvDielec3Dc} with system sizes
ranging up to $L = 50$. Here also, $\epsilon^{-1}$ decreases from
unity to zero, but the downward drift in the temperature at which
$\epsilon^{-1}$ deviates from unity seems to be even stronger than
for the 3D LG model. Additionally, the decrease towards zero does not
sharpen significantly with increasing $L$.

\begin{figure}[htbp]
  \centering
  \scalebox{0.8}{
    \hspace{-0.5cm}
    \includegraphics{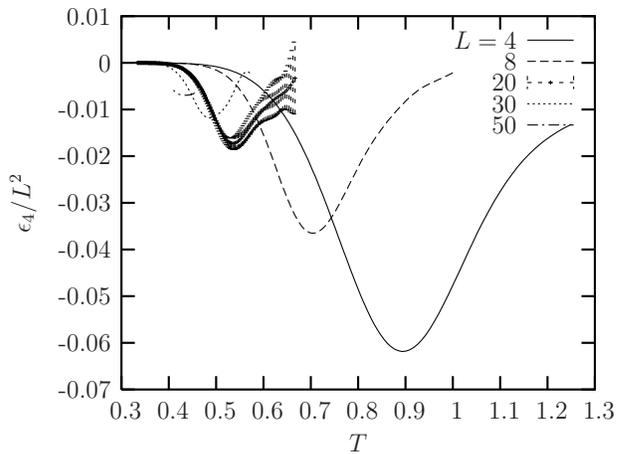}
  }
  \caption{The coefficient $\epsilon_4$ of the fourth order term of
    the expansion of the free energy, for the 3D CG. The depths
    decrease with increasing $L$ and seem to vanish as $L \rightarrow
    \infty$. Errorbars are shown for one of the systems for
    demonstration.}
  \label{fig:FourthOrder3Dc}
\end{figure}

\begin{figure}[htbp]
  \centering \scalebox{0.8}{ 
    \hspace{-2cm}
    \includegraphics{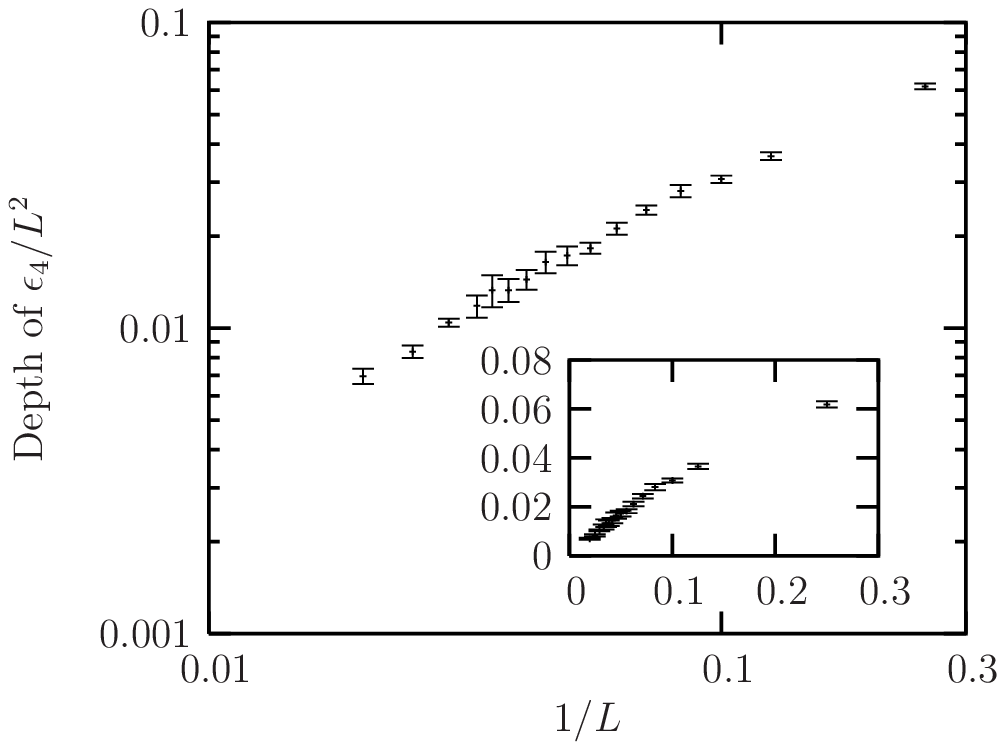} }
  \caption{Depth of the dip in the fourth order term shown in Figure
    \ref{fig:FourthOrder3Dc} for the 3D CG. The data are obtained from
    simulations of system sizes ranging from $L = 4$ to $L = 50$ and
    plotted both on a linear scale (inset) and on a log-log scale. It
    is clear that the dip vanishes in the thermodynamic limit.}
  \label{fig:Depth3Dc}
\end{figure}

We find a similar minimum in the fourth order term in the expansion of
the free energy for the 3D CG, $\epsilon_4/L^2$, shown in Fig.
\ref{fig:FourthOrder3Dc}. However, the dip
vanishes as $L \rightarrow \infty$ in the current model. This is
clearly shown in Fig. \ref{fig:Depth3Dc} in contrast to the Figs.
\ref{fig:Depth2D} and \ref{fig:Depth3D} of the other two models.

\begin{figure}[htbp]
  \centering \scalebox{0.8}{ 
    \hspace{-2cm}
    \includegraphics{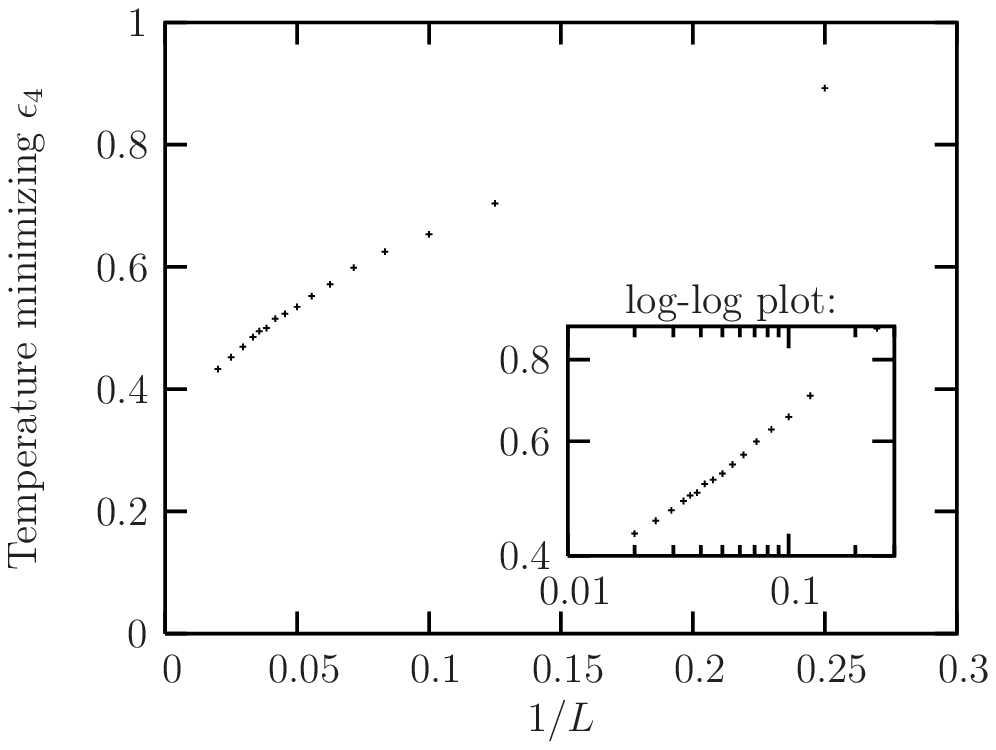} }
  \caption{Temperature minimizing $\epsilon_4$ as a function of
    inverse system size for the 3D CG system. The values are plotted
    both on a linear scale and on a log-log scale (inset).}
  \label{fig:Tc3Dc}
\end{figure}

For completeness we have included in Fig. \ref{fig:Tc3Dc} a plot of
the temperature locating the minimum in $\epsilon_4$ as a function of
system size also for the 3D CG. There is no phase transition
to which this temperature is associated, and the stronger
downward drift mentioned above is evident when contrasting this plot
to Fig. \ref{fig:Tc3D} of the 3D LG. The temperature is reduced by a 
factor 2 in the
largest system considered in the 3D CG compared to the smallest
whereas the variation is much smaller in the 3D LG. However,
there {\it is} a weak curvature in the log-log version of Fig. \ref{fig:Tc3Dc}. Performing
a similar extrapolation as we did for the other two models we end up
with a ``critical'' temperature $T_c = 0.24 \pm 0.04$.

\subsection{\label{subsec:ladning} Charge density}
\begin{figure}[htbp]
  \centering \scalebox{0.8}{ 
    \hspace{-2cm}
    \includegraphics{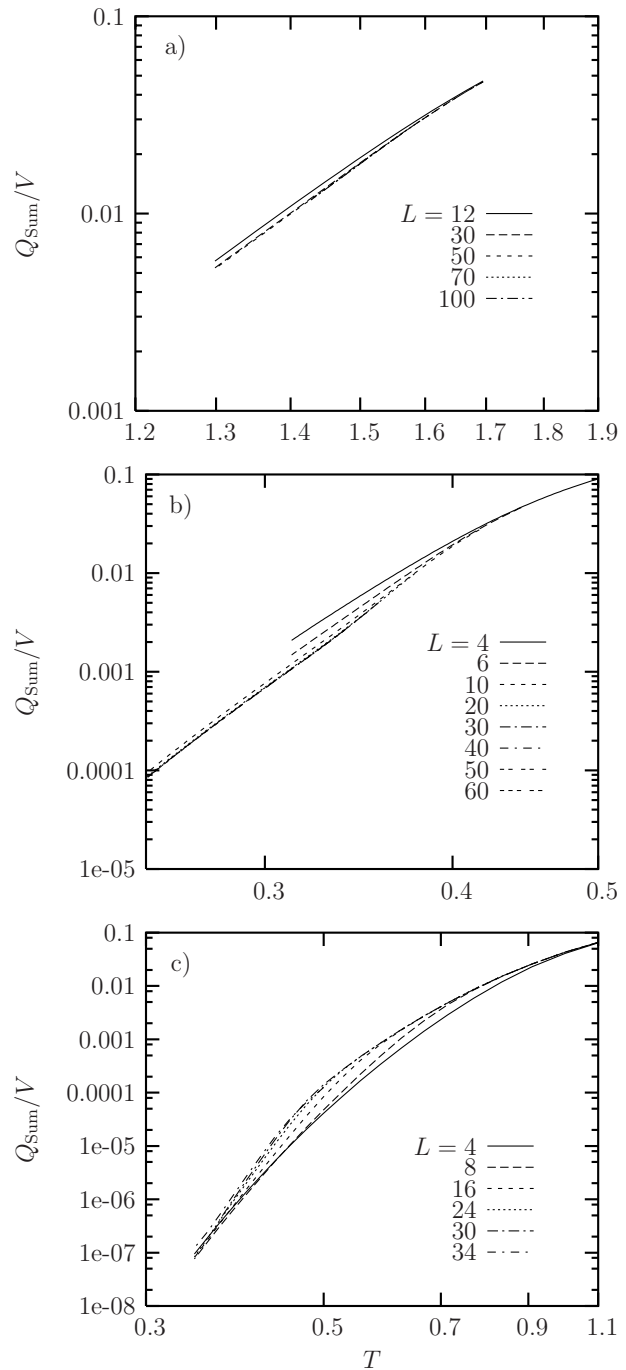} }
  \caption{Charge density $Q_{\mathrm{Sum}}/V$ plotted vs. temperature
    on log-log scales for the a) 2D CG, b) 3D LG and c) 3D CG models.
    The volume $V$ corresponds to the total number of sites $L^d$.
    Note that $Q_{\mathrm{Sum}}/V$ is independent of system size $L$
    in all three cases.}
  \label{fig:QSumMultiplot}
\end{figure}

Finally we present in Fig. \ref{fig:QSumMultiplot} the charge density
for the three models considered. In all three cases the charge
densities are independent of $L$ and from these curves we can
approximate the average separation $r_{\mathrm{mean}}$ between the
charges assuming uniform distribution,
\begin{equation}
  \label{eq:rmean}
  r_{\mathrm{mean}} = \left(\frac{1}{Q_{\mathrm{Sum}}/V}\right)^{1/d},
\end{equation}
where $d$ is the dimension. We consentrate on the ($L$-dependent)
temperatures which minimize $\epsilon_4$. In the two logarithmically
interacting models, $r_{\mathrm{mean}}$ ranges from $\sim 4$ for the
smallest systems and up to $\sim 8$ for the largest. In the 3D Coulomb
gas on the other hand, $r_{\mathrm{mean}}$ remains close to $L$ even
for the largest system sizes meaning that the systems are close to
their vacuum states at these temperatures. This strongly suggests that
the features we investigate are only extreme low-density effects in
the 3D CG model. Screening, which should take place at all temperatures in a system always being in a
metallic state, is not possible in this limit.

In the 2D CG and 3D LG models the situation is different. The
interesting temperature domains are smaller and the charge densities
are kept close to constant which in turn allows screening for the
largest systems.

\section{\label{sec:univers} COMMENTS ON UNIVERSALITY}

In the 2D CG, the universal jump to zero of the inverse dielectric constant
$\epsilon^{-1}$ is given by \cite{kosterlitz,minnhagen2}
\begin{equation}
  \epsilon^{-1} = \frac{2 T_c}{\pi}.
\label{univ_ratio}
\end{equation}
Using the estimate for the critical temperature found in
section \ref{subsec:sim2DCD}, the value at $T_c$ should, according to Eq.
(\ref{univ_ratio}), be $\epsilon^{-1} = 0.86 \pm 0.03$. This is in
agreement with Fig. \ref{fig:InvDielec2D}, since it is in this region the curves seem to split.

In Ref. \onlinecite{minnhagen}, it was speculated that the finite
negative value of the fourth order modulus $\langle \Upsilon_4 \rangle
\approx -0.130$ could be associated with a universal number. In the 2D CG,
$V_\bo{k} \sim L^2$ and $Q^y_{\bo{k}} \sim 2 \pi/L$ for large $L$, such that the
modification $\Delta \rightarrow \Delta/(\sqrt{2} \pi)$ turns
\eqref{eq:finalexpr} into
\begin{equation}
  \label{eq:modfinalexpr}
  F(\bo{T}) - F(0) = \frac{\Delta^2}{2} \epsilon^{-1} + \frac{\Delta^4}{4 !} \, 3 \epsilon_4.
\end{equation}
This means that if $\epsilon^{-1}$ corresponds to the helicity modulus
$\langle \Upsilon \rangle$, it is $3 \epsilon_4$ that corresponds to
the fourth order modulus $\langle \Upsilon_4 \rangle$. It is
interesting to notice that $3 \epsilon_4 = -0.141 \pm 0.015$ fits
nicely with the value found in Ref. \onlinecite{minnhagen}, speculated
to be a universal number. One may therefore speculate that the value
of $\epsilon_4$ at $T_c$ is a universal number independent of $T_c$.
Whether this is a sign of a true universality or a mere coincidence
requires further investigation.

One should also note that with this modification of $\Delta$, the
additional twist term in the XY-Hamiltonian \eqref{eq:ham2DXY} becomes
$\sqrt{2} \Delta \sin(2\pi y/L)/L$. It seems natural to suggest that
the net effect of a sine twist is given by its RMS-value, {\it i.e.}
$\big(1/L \int_0^L \sin^2(2\pi y/L) \du y\big)^{1/2} = 1/\sqrt{2}$.
This gives a net twist of $\Delta$ across the system, which is the
same as in Ref. \onlinecite{minnhagen}.

The universal jump of $\epsilon^{-1}$ in the 3D LG is given by the flow equations derived in Ref. \onlinecite{sudbo}. In our units, this jump is predicted to be
\begin{equation}
  \label{eq:hopp3DLG}
  \epsilon^{-1} = \frac{5 T_c}{2},
\end{equation}
and by using the critical temperature found in section \ref{subsec:sim3DLG}, this amounts to an $\epsilon^{-1}$ in the interval $(0.65,0.85)$. Since the different curves in Fig. \ref{fig:InvDielec3D} do not merge in the low-temperature regime, as they do in the 2D CG case, it is difficult to make a precise determination of the jump in the 3D LG based on these simulations. However, one can not rule out that the jump lies inside the interval mentioned.

\section{\label{sec:conclude} CONCLUDING REMARKS}

In this paper we have considered various quantitities related to a
possible phase transition in systems of point charges interacting with
bare logarithmic pair potentials, in 2D and 3D.  We have also carried
out comparisons with the results obtained in the 3D Coulomb gas in
some cases. The quantities we have focused on are the fluctuations of
the dipole moment, $\langle s^2 \rangle$, and the fourth order
coefficient of the free energy expanded in an appropriate twist.  We
have shown that the dipole moment fluctuations, associated with the
polarizability of the charge systems, has a scaling exponent
$\alpha(T)$ defined by $\langle s^2 \rangle \sim L^{\alpha(T)}$ which is
positive above some temperature and zero below this temperature for
the 2D CG and the 3D LG cases, and is an increasing function of
temperature. On the other hand, for the 3D CG case $\alpha(T)$ is
finite positive for all temperatures we have considered, and is a
{\it decreasing} function of temperature. This in itself strongly
suggests that the 3D LG has statistical physics much more akin to the
2D CG than to the 3D CG.  For the 2D CG we have demonstrated
that the inverse dielectric constant experiences a discontinuous jump
to zero at the phase transition. This has been
done by investigation of a series expansion of free energy using Monte
Carlo simulations. The possibility of a universal value of the fourth order
term proposed in Ref. \onlinecite{minnhagen} has also been commented
on, and a possible agreement with this value has been observed. 
The method developed in this paper will apply to any gas of
vortex loops or point charges with any interaction potential.
We have applied it to the 3D LG. Although it
would have been desirable to be able to access larger system sizes
than what we have been able to in the present paper, the results we
obtain for the 3D LG suggest that this model may also undergo a
metal-insulator transition with a discontinuity in the inverse
dielectric function at the critical point, in agreement
with the renormalization group results of Ref. \onlinecite{sudbo}.
 
\begin{acknowledgements}
  This work was supported by the Research Council of Norway, Grant
  Nos.  158518/431, 158547/431 (NANOMAT), and by the Norwegian High
  Performance Computing Consortium (NOTUR). One of us (S. K.)
  acknowledges support from the Norwegian University of Science and
  Technology. We acknowledge useful discussions with F. S. Nogueira and 
  Z. Tesanovic.
\end{acknowledgements}

\appendix
\section{\label{sec:appdual} DUALITY TRANSFORMATION}
The partition function for the XY-model with coupling constant $J=1$
is
\begin{equation}
  \label{eq:partXY}
  Z = \Pi_i \int \frac{\du \theta_i}{2 \pi} \, \ex{\beta \sr \cos (\nabla \theta - 2 \pi \bo{T})},
\end{equation}
where the sum is over all links between lattice points, $\nabla \theta
\equiv \theta_i - \theta_j$ and $\bo{T} (\bo{r})$ is the twist between
the two lattice points sharing the link $\bo{r}$. We will consider
three spatial dimensions and comment on any differences in 2D.
Applying the Villain approximation, we get
\begin{equation}
  \label{eq:partVillain}
  Z = \int {\cal D} \theta \sum_{\{\bo{n}\}} \ex{-\frac{\beta}{2} \sr (\nabla \theta - 2 \pi \bo{T} - 2 \pi \bo{n})^2}.
\end{equation}
$\bo{n} (\bo{r})$ is an integer-valued field taking care of the
periodicity of the cosine. By a Hubbard-Stratonovich decoupling, one
finds
\begin{equation}
  \label{eq:hubstrat}
  Z = \int {\cal D} \theta {\cal D} \bo{v} \sum_{\{\bo{n}\}} \ex{- \sr [\frac{1}{2 \beta} \bo{v}^2 + \im \bo{v} \cdot (\nabla \theta - 2 \pi \bo{T} - 2 \pi \bo{n})]}.
\end{equation}
The summation over $\bo{n}$ may now be evaluated using the Poisson
summation formula,
\begin{equation}
  \label{eq:poissondef}
  \sum_{n=-\infty}^{\infty} \ex{2 \pi \im \, n v} = \sum_{l=-\infty}^{\infty} \delta (v-l),
\end{equation}
at each dual lattice point, yielding
\begin{equation}
  \label{eq:poiss1}
  Z = \int {\cal D} \theta \sum_{\{\bo{l}\}} \ex{ \sr 2 \pi \im \, \bo{l} \cdot \bo{T} - \im \, \bo{l} \cdot \nabla \theta - \frac{1}{2 \beta} \bo{l}^2}.
\end{equation}
The field $\bo{l} (\bo{r})$ is integer-valued. Now, performing a
partial summation on the second term in the exponent, the
$\theta$-integration may be carried out. This produces the constraint
that $\bo{l}$ must be divergence-free, solved by the introduction of
another integer-valued field such that $\bo{l} = \nabla \times
\bo{h}$. Note that $\bo{h} (\bo{r})$ is a scalar in 2D. The partition
function is now
\begin{equation}
  \label{eq:thetagone}
  Z = \sum_{\{\bo{h}\}} \ex{ \sr 2 \pi \im \, (\nabla \times \bo{h}) \cdot \bo{T} - \frac{1}{2 \beta} (\nabla \times \bo{h})^2 },
\end{equation}
and we observe that $\bo{h} \rightarrow \bo{h} + \nabla \phi$ is a
gauge transformation. In two dimensions, the corresponding gauge
transformation is $h \rightarrow h + c$, where $c$ is a constant.
Using Poisson's summation formula once more, we get
\begin{equation}
  \label{eq:poiss2}
  Z = \int {\cal D} \bo{h} \sum_{\{\bo{m}\}} \ex{ \sr 2 \pi \im \, (\nabla \times \bo{h}) \cdot \bo{T} - \frac{1}{2 \beta} (\nabla \times \bo{h})^2 + 2 \pi \im \, \bo{h} \cdot \bo{m}},
\end{equation}
leaving $\bo{h}$ no longer integer-valued. The field $\bo{m} (\bo{r})$
is what corresponds to vortex excitations in the XY model. The gauge
invariance of the theory produces the constraint $\sr \phi \, (\nabla
\cdot \bo{m})=0$ for all configurations of $\bo{m}$. Choosing for
instance $\phi = \nabla \cdot \bo{m}$, it is clear that $\bo{m}$ must
be divergence-free, {\it i.e.} the field lines are closed loops. In
2D, the corresponding constraint is $\sum_{\bo{r}} m = 0$, indicating
an overall charge neutrality in the 2D Coulomb gas or zero total
vorticity in the 2D XY model.

By another partial summation, we are now left with a Maxwell term and
a coupling term between the gauge field $\bo{h}$ and the current
$\bo{M} (\bo{r}) \equiv \bo{m} + \nabla \times \bo{T}$:
\begin{equation}
  \label{eq:maxwell+coupling}
  Z = \int {\cal D} \bo{h} \sum_{\{\bo{m}\}} \ex{ \sr 2 \pi \im \, \bo{h} \cdot \bo{M} - \frac{1}{2 \beta} (\nabla \times \bo{h})^2 }.
\end{equation}
One may now perform a partial integration in the second term and use
the gauge where $\nabla \cdot \bo{h} = 0$, such that \linebreak
$\nabla \times \nabla \times \bo{h} = - \nabla^2 \bo{h}$. Then, by
going to Fourier space and completing squares, the
$\bo{h}$-integration becomes Gaussian. This leaves us with
\begin{equation}
  \label{eq:hintout}
  Z = Z_0 \sum_{\{\bo{m}\}} \ex{ \frac{2 \beta \pi^2}{N} \sum_{\bo{q}} \bo{M}_{\bo{q}} G^{-1}_{\bo{q}} \bo{M}_{-\bo{q}} },
\end{equation} 
where $\nabla^2 \ex{\pm \im \bo{q} \bo{r}} \equiv \ex{\pm \im \bo{q}
  \bo{r}} G_{\bo{q}}$ and $Z_0$ is a constant. Defining the discrete
Laplacian by
\begin{equation}
  \label{eq:}
  \Delta^{2} f(\bo{r}) = \sum_{\mu} \left[f(\bo{r}+\hat{e}_\mu) + f(\bo{r}-\hat{e}_\mu) - 2 f(\bo{r})\right],
\end{equation}
it is clear that $G_{\bo{q}} = -2 \left( d-\sum_{\mu=1}^d \cos
  q_{\mu}\right)$, denoting the number of space dimensions by $d$.
Returning to real-space representation, we arrive at
\begin{equation}
  \label{eq:sluttpart}
  Z = Z_0 \sum_{\{\bo{m}\}} \ex{ - \frac{\beta}{2} \sum_{\bo{r}_i, \bo{r}_j} \bo{M}(\bo{r}_i) \, V(|\bo{r}_i - \bo{r}_j|) \, \bo{M}(\bo{r}_j) },
\end{equation}
the interaction being given by
\begin{equation}
  \label{eq:Vslutt}
  V(\bo{r}) = \frac{2 \pi^2}{L^2} \sum_{\bo{q}} \frac{\ex{\im \bo{q} \cdot \bo{r}}}{d - \sum_{\mu=1}^{d} \cos q_\mu}.
\end{equation}

\section{\label{sec:appTaylor} EXPANSION OF FREE ENERGY}

Consider the Hamiltonian
\begin{equation}
H_0 = \frac{1}{2} \sum_{i,j} \bo{m}_i V_{ij} \bo{m}_j ,
\label{eq:ham0}
\end{equation}
describing a 3D system of integer-valued currents $\bo{m}$ on a
lattice interacting via the potential $V_{ij} = V(|\bo{r}_i -
\bo{r}_j|)$. We impose periodic boundary conditions on the system.
Perturbing the field $\bo{m}$ with a transversal twist turns
\eqref{eq:ham0} into
\begin{equation}
H = \frac{1}{2} \sum_{i,j} (\bo{m} + \nabla \times \bo{T})_i V_{ij} (\bo{m} + \nabla \times \bo{T})_j.
\label{eq:ham}
\end{equation}
We let the linear system size be $L$ and define the discrete Fourier
transform by
\begin{equation}
f_{\bo{q}} = \sum_{\bo{r}} f(\bo{r}) \, \ex{\im \bo{q} \cdot \bo{r}},
\label{eq:deffour}
\end{equation}
where $\bo{r}=(n_x,n_y,n_z)$ and $n_i = 0,...,L-1$. The inverse
transform is
\begin{equation}
f(\bo{r})  = \frac{1}{N} \sum_{\bo{q}} f_{\bo{q}} \, \ex{-\im \bo{q} \cdot \bo{r}},
\label{eq:definvfour}
\end{equation}
where $\bo{q} = \frac{2 \pi}{L} (k_x,k_y,k_z)$ and $k_i =
-L/2+1,...,L/2$. $N$ is the number of lattice sites. Let us also
define $Q^{\nu}_{\pm \bo{q}}$ by $\Delta^{\nu} \ex{\pm \im \bo{q}
  \cdot \bo{r}} = \ex{\pm \im \bo{q} \cdot \bo{r}} Q^{\nu}_{\pm
  \bo{q}}$, where $\Delta^{\nu}$ is a lattice derivative. In Fourier
representation the Hamiltonian becomes
\begin{widetext}
\begin{equation}
H = \frac{1}{2N} \sum_{\bo{q}} \left(m^{\mu}_{\bo{q}} + \varepsilon^{\mu \nu \lambda} Q^{\nu}_{-\bo{q}} T^{\lambda}_{\bo{q}} \right)V_{\bo{q}} \left(m^{\mu}_{-\bo{q}} + \varepsilon^{\mu \rho \sigma} Q^{\rho}_{\bo{q}} T^{\sigma}_{-\bo{q}} \right).
\label{eq:hamfour}
\end{equation}
For later use, we calculate the derivative of $H$, which is
\begin{equation}
\delH{\alpha}{1} = \frac{1}{N} \varepsilon^{\mu \nu \alpha} Q^{\nu}_{-\bo{q}_1} (m^{\mu}_{-\bo{q}_1} + \varepsilon^{\mu \rho \sigma} Q^{\rho}_{\bo{q}_1} T^{\sigma}_{-\bo{q}_1}) V_{\bo{q}_1}.
\label{eq:delH}
\end{equation}
We also note that 
\begin{equation}
\frac{\partial^2 H}{\partial T^{\alpha}_{\bo{q}_1} \partial T^{\beta}_{\bo{q}_2}} = \frac{1}{N} \varepsilon^{\mu \nu \alpha} \varepsilon^{\mu \rho \beta} Q^{\nu}_{-\bo{q}_1} Q^{\rho}_{\bo{q}_1} V_{\bo{q}_1} \delta_{\bo{q}_1 + \bo{q}_2,0}   
\label{eq:del2H}
\end{equation}
is independent of $\bo{m}$ and that all higher order derivatives are
zero.

The free energy is given by $F = - T \ln Z$, where the partition
function is
\begin{equation}
  \label{eq:partfunc}
  Z = \sm \ex{-H/T},
\end{equation}
summing over all possible configurations of $\bo{m}$. By Taylor
expansion of the free energy in the twist, we get
\begin{equation}
  \label{eq:taylorrealspace}
  F(\bo{T})-F(0) = \sum_{\alpha} \sum_{\bo{r}_1} \frac{\partial F}{\partial T^\alpha(\bo{r}_1)}\bigg|_{\bo{T} = 0} T^\alpha(\bo{r}_1) + \sum_{\alpha, \beta} \sum_{\bo{r}_1 , \bo{r}_2} \frac{\partial^2 F}{\partial T^\alpha(\bo{r}_1) \partial T^\beta(\bo{r}_2)}\bigg|_{\bo{T} = 0} T^\alpha(\bo{r}_1) T^\beta(\bo{r}_2) + ...
\end{equation}
Note that $F(\bo{T} = 0)$ refers to the free energy of the unperturbed
system described by $H_0$. By writing each term in the series in
Fourier representation, one finds the equivalent expansion in Fourier
components of the twist, {\it i.e.}
\begin{equation}
  \label{eq:freeseries}
  F(\bo{T}) - F(0) = \sum_{\alpha} \sum_{\bo{q}_1} \frac{\partial F}{\partial T^{\alpha}_{\bo{q}_1}}\bigg|_{\bo{T} = 0} T^{\alpha}_{\bo{q}_1} + \sum_{\alpha, \beta} \sum_{\bo{q}_1 \bo{q}_2} \frac{\partial^2 F}{\partial T^{\alpha}_{\bo{q}_1} \partial T^{\beta}_{\bo{q}_2}}\bigg|_{\bo{T} = 0} \frac{T^{\alpha}_{\bo{q}_1} T^{\beta}_{\bo{q}_2}}{2} + ... 
\end{equation}
The first derivative becomes
\begin{equation}
  \label{eq:1deriv}
  \frac{\partial F}{\partial T^{\alpha}_{\bo{q}_1}} = \frac{1}{Z} \sm \delH{\alpha}{1} \bz \equiv \left\langle \delH{\alpha}{1} \right\rangle.
\end{equation}
Proceeding, we find 
\begin{equation}
  \label{eq:2deriv}
  \frac{\partial^2 F}{\partial T^{\alpha}_{\bo{q}_1} \partial T^{\beta}_{\bo{q}_2}} = \frac{1}{T} \fv{\delH{\alpha}{1}} \fv{\delH{\beta}{2}} + \ddelH{\alpha}{1}{\beta}{2} - \frac{1}{T} \fv{\delH{\alpha}{1}\delH{\beta}{2}}
\end{equation}
for the second derivative and
\begin{equation}
  \label{eq:3deriv}
  \begin{split}
  \frac{\partial^3 F}{\partial T^{\alpha}_{\bo{q}_1} \partial T^{\beta}_{\bo{q}_2} \partial T^{\gamma}_{\bo{q}_3}} & = \frac{1}{T^2} \left[2 \fv{\delH{\alpha}{1}} \fv{\delH{\beta}{2}} \fv{\delH{\gamma}{3}} + \fv{\delH{\alpha}{1}\delH{\beta}{2}\delH{\gamma}{3}} \right.\\ 
& \qquad \quad - \fv{\delH{\alpha}{1}} \fv{\delH{\beta}{2}\delH{\gamma}{3}}  - \fv{\delH{\beta}{2}} \fv{\delH{\alpha}{1}\delH{\gamma}{3}} \\ 
& \left. \qquad \quad - \fv{\delH{\gamma}{3}} \fv{\delH{\alpha}{1}\delH{\beta}{2}} \right]
\raisetag{1cm}
\end{split}
\end{equation}
for the third. We have exploited the fact that third derivatives of
$H$ vanishes. The fourth derivative is found to be
\begin{equation}
  \label{eq:4deriv}
  \begin{split}
  \frac{\partial^4 F}{\partial T^{\alpha}_{\bo{q}_1} \partial T^{\beta}_{\bo{q}_2} \partial T^{\gamma}_{\bo{q}_3} \partial T^{\delta}_{\bo{q}_4}} & = \frac{1}{T^3} \left\{ 6 \fv{\delH{\alpha}{1}} \fv{\delH{\beta}{2}} \fv{\delH{\gamma}{3}} \fv{\delH{\delta}{4}}  \right. \\ 
& \qquad \quad - 2 \left[ \fv{\delH{\alpha}{1}} \fv{\delH{\beta}{2}} \fv{\delH{\gamma}{3}\delH{\delta}{4}} + \fv{\delH{\alpha}{1}} \fv{\delH{\gamma}{3}} \fv{\delH{\beta}{2}\delH{\delta}{4}} \right.  \\ 
& \qquad \quad \quad + \fv{\delH{\alpha}{1}} \fv{\delH{\delta}{4}} \fv{\delH{\beta}{2}\delH{\gamma}{3}} + \fv{\delH{\beta}{2}} \fv{\delH{\gamma}{3}} \fv{\delH{\alpha}{1}\delH{\delta}{4}} \\ 
& \left. \qquad \quad \quad  + \fv{\delH{\beta}{2}} \fv{\delH{\delta}{4}} \fv{\delH{\alpha}{1}\delH{\gamma}{3}} + \fv{\delH{\gamma}{3}} \fv{\delH{\delta}{4}} \fv{\delH{\alpha}{1}\delH{\beta}{2}} \right] \\ 
& \qquad \quad + \fv{\delH{\alpha}{1}} \fv{\delH{\beta}{2} \delH{\gamma}{3} \delH{\delta}{4}} + \fv{\delH{\beta}{2}} \fv{\delH{\alpha}{1} \delH{\gamma}{3} \delH{\delta}{4}} \\ 
& \qquad \quad  + \fv{\delH{\gamma}{3}} \fv{\delH{\alpha}{1} \delH{\beta}{2} \delH{\delta}{4}} + \fv{\delH{\delta}{4}} \fv{\delH{\alpha}{1} \delH{\beta}{2} \delH{\gamma}{3}} \\
& \qquad \quad + \fv{\delH{\alpha}{1} \delH{\beta}{2}} \fv{\delH{\gamma}{3} \delH{\delta}{4}} + \fv{\delH{\alpha}{1} \delH{\gamma}{3}} \fv{\delH{\beta}{2} \delH{\delta}{4}}  \\ 
& \left. \qquad \quad  + \fv{\delH{\alpha}{1} \delH{\delta}{4}} \fv{\delH{\beta}{2} \delH{\gamma}{3}} - \fv{\delH{\alpha}{1}\delH{\beta}{2}\delH{\gamma}{3}\delH{\delta}{4}} \right\}.
\raisetag{1cm}
\end{split}
\end{equation}
Remembering that 
\begin{equation}
  \label{eq:delHT0}
  \delH{\alpha}{1}\bigg|_{\bo{T}=0} = \frac{1}{N} \varepsilon^{\mu \nu \alpha} Q^{\nu}_{-\bo{q}_1} m^\mu_{-\bo{q}_1} V_{\bo{q}_1},
\end{equation}
it is straightforward to write the derivatives at zero twist as
$\bo{m}$-correlators. However, in many cases these expressions may be
simplified further. If the sum over all possible configurations
$\{\bo{m}\}$ is symmetric around zero, one finds that all odd-order
correlators are zero, resulting in
\begin{equation}
  \label{eq:1&3zero}
  \frac{\partial F}{\partial T^{\alpha}_{\bo{q}_1}}\bigg|_{\bo{T}=0} = \frac{\partial^3 F}{\partial T^{\alpha}_{\bo{q}_1} \partial T^{\beta}_{\bo{q}_2} \partial T^{\gamma}_{\bo{q}_3}}\bigg|_{\bo{T}=0} = 0.
\end{equation} 
Furthermore, since $V_{ij} = V(|\bo{r}_i - \bo{r}_j|)$, {\it i.e.} we
have a translationally invariant system, the even-order correlators
are subject to relations like
\begin{equation}
  \label{eq:2corr}
  \fv{m^{\mu}_{-\bo{q}_1} m^{\nu}_{-\bo{q}_2}} = \fv{m^{\mu}_{-\bo{q}_1} m^{\nu}_{-\bo{q}_2}} \delta_{\bo{q}_1 + \bo{q}_2, 0}
\end{equation}
and
\begin{equation}
  \label{eq:4corr}
  \fv{m^{\mu}_{-\bo{q}_1} m^{\nu}_{-\bo{q}_2} m^{\kappa}_{-\bo{q}_3} m^{\lambda}_{-\bo{q}_4}} = \fv{m^{\mu}_{-\bo{q}_1} m^{\nu}_{-\bo{q}_2} m^{\kappa}_{-\bo{q}_3} m^{\lambda}_{-\bo{q}_4}} \delta_{\bo{q}_1 + \bo{q}_2 + \bo{q}_3 + \bo{q}_4,0}.
\end{equation}
Thus, we find
\begin{equation}
  \label{eq:2derivzero}
  \frac{\partial^2 F}{\partial T^{\alpha}_{\bo{q}_1} \partial T^{\beta}_{\bo{q}_2}}\bigg|_{\bo{T}=0} = \frac{\varepsilon^{\mu \sigma \alpha} \varepsilon^{\nu \rho \beta} Q^\sigma_{-\bo{q}_1} Q^\rho_{-\bo{q}_2} V_{\bo{q}_1} \delta_{\bo{q}_1 + \bo{q}_2, 0}}{N} \left(\delta^{\mu \nu} - \frac{V_{\bo{q}_2}}{N T} \fv{m^\mu_{-\bo{q}_1} m^\nu_{-\bo{q}_2}} \right)
\end{equation}
for the second derivative and 
\begin{equation}
  \label{eq:4derivzero}
  \begin{split}
  \frac{\partial^4 F}{\partial T^{\alpha}_{\bo{q}_1} \partial T^{\beta}_{\bo{q}_2} \partial T^{\gamma}_{\bo{q}_3} \partial T^{\delta}_{\bo{q}_4}}\bigg|_{\bo{T}=0} & = \frac{\varepsilon^{\mu \sigma \alpha} \varepsilon^{\nu \rho \beta} \varepsilon^{\kappa \tau \gamma} \varepsilon^{\lambda \eta \delta} Q^\sigma_{-\bo{q}_1} Q^\rho_{-\bo{q}_2} Q^\tau_{-\bo{q}_3} Q^\eta_{-\bo{q}_4} V_{\bo{q}_1} V_{\bo{q}_2} V_{\bo{q}_3}  V_{\bo{q}_4}}{N^4 T^3} \\
& \quad \times \left\{ \fv{m^\mu_{-\bo{q}_1} m^\nu_{-\bo{q}_2}} \fv{m^\kappa_{-\bo{q}_3} m^\lambda_{-\bo{q}_4}} \delta_{\bo{q}_1 + \bo{q}_2, 0} \, \delta_{\bo{q}_3 + \bo{q}_4, 0} \right.\\
& \qquad + \fv{m^\mu_{-\bo{q}_1} m^\kappa_{-\bo{q}_3}} \fv{m^\nu_{-\bo{q}_2} m^\lambda_{-\bo{q}_4}} \delta_{\bo{q}_1 + \bo{q}_3, 0} \, \delta_{\bo{q}_2 + \bo{q}_4, 0} \\
& \qquad +  \fv{m^\mu_{-\bo{q}_1} m^\lambda_{-\bo{q}_4}} \fv{m^\nu_{-\bo{q}_2} m^\kappa_{-\bo{q}_3}} \delta_{\bo{q}_1 + \bo{q}_4, 0} \, \delta_{\bo{q}_2 + \bo{q}_3, 0} \\
& \left. \qquad - \fv{m^\mu_{-\bo{q}_1} m^\nu_{-\bo{q}_2} m^\kappa_{-\bo{q}_3} m^\lambda_{-\bo{q}_4}} \delta_{\bo{q}_1 + \bo{q}_2 + \bo{q}_3 + \bo{q}_4, 0}  \right\}
\raisetag{1cm}
\end{split}
\end{equation}
for the fourth. These expressions may also be applied to a gas of
point charges in 2D or 3D, that is when $m$ is a scalar field. One way
to do this is by replacing $\nabla \times \bo{T}$ in \eqref{eq:ham}
with its $z$-component $\varepsilon^{z \nu \lambda}\Delta^\nu
T^\lambda$, with the consequence that the greek letter summations may
be taken over $x$ and $y$ only. For the second derivative, this
results in
\begin{equation}
  \label{eq:2derivpoint}
  \frac{\partial^2 F}{\partial T^{\alpha}_{\bo{q}_1} \partial T^{\beta}_{\bo{q}_2}}\bigg|_{\bo{T}=0} = \frac{\varepsilon^{z \sigma \alpha} \varepsilon^{z \rho \beta} Q^\sigma_{-\bo{q}_1} Q^\rho_{\bo{q}_1} V_{\bo{q}_1} \delta_{\bo{q}_1 + \bo{q}_2, 0}}{N} \left(1 - \frac{V_{\bo{q}_1}}{N T} \fv{m_{\bo{q}_1} m_{-\bo{q}_1}} \right),
\end{equation}
\end{widetext}
where we have applied $V_{\bo{q}} = V_{-\bo{q}}$. We recognize the
paranthesis as the Fourier transform of the inverse dielectric
response function $\epsilon^{-1}(\bo{q_1})$ in the low density limit.
Note that the factor
\begin{equation}
  \label{eq:projop}
\varepsilon^{z \sigma \alpha} \varepsilon^{z \rho \beta} Q^\sigma_{-\bo{q}_1} Q^\rho_{\bo{q}_1} = Q^\sigma_{\bo{q}_1} Q^\sigma_{-\bo{q}_1} \left(1-\frac{Q^\alpha_{\bo{q}_1} Q^\beta_{-\bo{q}_1}}{Q^\sigma_{\bo{q}_1} Q^\sigma_{-\bo{q}_1}} \right)
\end{equation}
is a projection operator times $Q^\sigma_{\bo{q_1}}
Q^\sigma_{-\bo{q_1}} \sim q_{1x}^2 + q_{1y}^2$, reflecting the
transversality of the twist.

To arrive at eq. \eqref{eq:finalexpr}, we chose the twist
\eqref{eq:twistchoice} and computed the sums appearing in the
expansion \eqref{eq:freeseries} for both the second and fourth order
term. The sum over direction is trivial, since our twist points in the
$x$-direction. The sum over momenta is also managable, since
$T^x_{\bo{q}}$ has nonzero values only for $\bo{q} = (0,\pm 2\pi/L)$.
This sum gives two contributions in the second order term, due to the
restriction $\delta_{\bo{q}_1+\bo{q}_2,0}$. The same argument results
in four contributions for the three terms in \eqref{eq:4derivzero}
being a product of two second order correlators. The term containing a
fourth order correlator will give six contributions due to the
restriction $\delta_{\bo{q}_1 + \bo{q}_2 + \bo{q}_3 + \bo{q}_4, 0}$.

\section{\label{sec:app6order} HIGHER ORDER TERMS}

Using the method described in this paper involves extrapolation to $L
\rightarrow \infty$ and deciding whether or not the fourth order term
in the expansion \eqref{eq:freeseries} goes to zero or to a finite
nonzero value. This procedure could in some cases be difficult.
However, if the fourth order term had turned out to be zero in the
thermodynamic limit, it would {\it not} necessarily mean that the
second order term, the inverse dielectric response function, would
have to go continuously to zero. In fact, if one were able to prove
that the fourth order term is negative {\it or} zero, one could go on
to investigate the sixth order term instead. If it then turned out
that the value of the sixth order term was hard to establish, one
could in principle repeat the procedure and go to higher order terms.
We therefore include the sixth derivative here. To simplify
calculations, we work with a twist in the $x$-direction only.
\begin{widetext}
\begin{equation}
  \label{eq:6deriv}
  \begin{split}
    & \frac{\partial^6 F}{\partial T^x_{\bo{q}_1} \partial T^x_{\bo{q}_2} \partial T^x_{\bo{q}_3} \partial T^x_{\bo{q}_4} \partial T^x_{\bo{q}_5} \partial T^x_{\bo{q}_6}} = \\ 
& \frac{1}{T^5} \left\{120 \fv{\Hx{1}}\fv{\Hx{2}}\fv{\Hx{3}} \fv{\Hx{4}}\left[\fv{\Hx{5}}\fv{\Hx{6}} - 3 \fv{\Hx{5} \Hx{6}}  \right] \right.\\ 
& + 18 \fv{\Hx{1} \Hx{2}}  \left[13 \fv{\Hx{3}}\fv{\Hx{4}}\fv{\Hx{5} \Hx{6}} - \fv{\Hx{3} \Hx{4}}\fv{\Hx{5} \Hx{6}} \right] \\
&  + 2 \fv{\Hx{1} \Hx{2} \Hx{3}} \left[5 \fv{\Hx{4} \Hx{5} \Hx{6}}  - 48 \fv{\Hx{4}} \fv{\Hx{5} \Hx{6}} + 60 \fv{\Hx{4}}\fv{\Hx{5}}\fv{\Hx{6}} \right]\\
& + 15  \fv{\Hx{1} \Hx{2} \Hx{3} \Hx{4}} \left[\fv{\Hx{5} \Hx{6}}  - 2 \fv{\Hx{5}}\fv{\Hx{6}} \right]   \\ 
& \left. + 6 \fv{\Hx{1}}\fv{\Hx{2} \Hx{3} \Hx{4} \Hx{5} \Hx{6}} - \fv{\Hx{1} \Hx{2} \Hx{3} \Hx{4} \Hx{5} \Hx{6}} \right\} \\
& - \frac{12}{T^4} \Hxx{1}{2} \left\{2  \fv{\Hx{3}} \fv{\Hx{4}} \fv{\Hx{5} \Hx{6}} + 2 \fv{\Hx{3} \Hx{4}} \fv{\Hx{5} \Hx{6}} \right. \\ 
& \left. + \fv{\Hx{3}} \fv{\Hx{4} \Hx{5} \Hx{6}} \right\} + \frac{12}{T^3} \Hxx{1}{2}\Hxx{3}{4} \left\{ \fv{\Hx{5} \Hx{6}} + 2 \fv{\Hx{5}}\fv{\Hx{6}}\right\} .
  \end{split}
  \raisetag{1cm}
\end{equation}
Note that we are allowed to permute the momenta
$\bo{q}_1,...,\bo{q}_6$, since these are summed over in the free
energy expansion. Assuming vanishing odd-order correlators and
imposing that $m$ is a scalar field gives
\begin{equation}
\label{eq:sixfinal}
\begin{split}
\frac{\partial^6 F}{\partial T^x_{\bo{q}_1} \partial T^x_{\bo{q}_2} \partial T^x_{\bo{q}_3} \partial T^x_{\bo{q}_4} \partial T^x_{\bo{q}_5} \partial T^x_{\bo{q}_6}} & = \frac{Q^y_{-\bo{q}_1} Q^y_{-\bo{q}_2} Q^y_{-\bo{q}_3} Q^y_{-\bo{q}_4} Q^y_{-\bo{q}_5} Q^y_{-\bo{q}_6} V_{\bo{q}_1} V_{\bo{q}_2} V_{\bo{q}_3}  V_{\bo{q}_4}}{N^4 T^3} \\
& \times \left\{ 12 \fv{m_{-\bo{q}_1} m_{-\bo{q}_2}} \left[1 - \frac{2 V_{\bo{q}_5}}{N T} \fv{m_{-\bo{q}_3} m_{-\bo{q}_4}}\right] \delta_{\bo{q}_1+\bo{q}_2,0} \, \delta_{\bo{q}_3+\bo{q}_4,0} \, \delta_{\bo{q}_5+\bo{q}_6,0}  \right.\\
&\quad -  \frac{V_{\bo{q}_5} V_{\bo{q}_6}}{N^2 T^2} \bigg[\fv{m_{-\bo{q}_1} m_{-\bo{q}_2} m_{-\bo{q}_3} m_{-\bo{q}_4} m_{-\bo{q}_5} m_{-\bo{q}_6}} \delta_{\bo{q}_1+\bo{q}_2 +\bo{q}_3+\bo{q}_4+\bo{q}_5+\bo{q}_6,0} \\
& \qquad \qquad \quad - 3 \fv{m_{-\bo{q}_1} m_{-\bo{q}_2}} \delta_{\bo{q}_1+\bo{q}_2,0} \bigg(5 \fv{m_{-\bo{q}_3} m_{-\bo{q}_4} m_{-\bo{q}_5} m_{-\bo{q}_6}} \delta_{\bo{q}_3+\bo{q}_4+\bo{q}_5+\bo{q}_6,0} \\
& \left. \hspace{4.3cm} - 6 \fv{m_{-\bo{q}_3} m_{-\bo{q}_4}} \fv{m_{-\bo{q}_5} m_{-\bo{q}_6}} \delta_{\bo{q}_3+\bo{q}_4,0} \, \delta_{\bo{q}_5+\bo{q}_6,0} \bigg) \bigg] \right\}.
\end{split}
\raisetag{1cm}
\end{equation}

\end{widetext}
\bibliography{2DCG}

\end{document}